 \documentclass[10pt, twocolumn]{IEEEtran}

\usepackage{amsmath}
\usepackage{amsmath,amsthm}
\usepackage{cite}   
\usepackage{graphicx}  
\usepackage{amsmath, amsopn, psfrag, amsthm}   
\usepackage{amssymb}
\usepackage{color}

\usepackage{algorithm}
\usepackage{amsmath}
\usepackage{algorithmic}
\usepackage[utf8]{inputenc}
\usepackage[english]{babel}
\newtheorem{theorem}{Theorem}
\newtheorem{lemma}{Lemma}
\newtheorem{remark}{Remark}

\usepackage{psfrag}
\usepackage{epsfig}
\graphicspath{{./img/}}
\DeclareGraphicsExtensions{.pdf,.jpeg,.png, .eps}
\DeclareMathOperator*{\argmax}{arg\, max}

\usepackage{amssymb,amsmath,mathtools}

\def\bomega{{\boldsymbol{\Omega}}}

\def\bPhi{{\boldsymbol{\Phi}}}

\usepackage{cite}   
\usepackage{graphicx}  
\usepackage{amsmath, amsopn}   
\usepackage{amssymb}
\usepackage{psfrag}
\usepackage{epsfig}
\usepackage{stfloats}
\usepackage{lipsum}
\usepackage{multicol}
\graphicspath{{./img/}}
\DeclareGraphicsExtensions{.pdf,.jpeg,.png}
\begin{document}
\vspace{-.2cm}
\title{On the Performance of Backhaul Constrained Cell-Free Massive MIMO with Linear Receivers}
\vspace{-.2cm}
\linespread{.9}
\author{
\vspace{-.1cm}
$\!\!\!$\IEEEauthorblockN{Manijeh Bashar\IEEEauthorrefmark{1}$\!$, Hien Quoc Ngo\IEEEauthorrefmark{2}$\!$, Alister G. Burr\IEEEauthorrefmark{1}$\!$, Dick Maryopi\IEEEauthorrefmark{1}$\!$, Kanapathippillai Cumanan\IEEEauthorrefmark{1}$\!\!$, and Erik G. Larsson\IEEEauthorrefmark{3}}$\!\!\!\!$\\
 \IEEEauthorblockA{\IEEEauthorrefmark{1}Department of Electronic Engineering, University of York, UK, \IEEEauthorblockA{\IEEEauthorrefmark{2}School of Electronics, Electrical Engineering and Computer Science, Queen's University Belfast, UK,}
\IEEEauthorblockA{\IEEEauthorrefmark{3}Department of Electrical Engineering (ISY), Linköping University, Sweden.} Email:{ \{mb1465, alister.burr, dm1110, kanapathippillai.cumanan\}@york.ac.uk}, hien.ngo@qub.ac.uk, erik.g.larsson@liu.se$\!\!\!\!\!\!$}
\vspace{-.3in}}
\vspace{-20in}%
\linespread{.9}
\maketitle
\vspace{-.1in}
\begin{abstract}
Limited-backhaul cell-free Massive multiple-input multiple-output (MIMO), in which the fog radio access network (F-RAN) is implemented to exchange the information between access points (APs) and the central processing unit (CPU), is investigated. We introduce a novel approach where the APs estimate the channel and send back the quantized version of the estimated channel and the quantized version of the received signal to the central processing unit. The Max algorithm and the Bussgang theorem are exploited to model the optimum uniform quantization. The ergodic achievable rates are derived. We show that exploiting microwave wireless backhaul links and using a small number of bits to quantize the estimated channel and the received signal, the performance of limited-backhaul cell-free Massive MIMO closely approaches the performance of cell-free Massive MIMO with perfect backhaul links.

\end{abstract}
\vspace{-0.22in}
\section{Introduction}
\vspace{-0.01in}
 \let\thefootnote\relax\footnotetext{The work of A. Burr and K. Cumanan was supported by H2020- MSCA-RISE-2015 under grant number 690750.}
In this paper, similar to the methology in \cite{ourvtc18,ourjournal2,ouricc2}, 
we combine cell-free Massive multiple-input multiple-output (MIMO) systems with fog radio access network (F-RAN). Moreover, we study the effect of limited-capacity links from the APs to the central processing unit (CPU) (or from the remote radio heads (RRHs) to the base band unit (BBU)). The limited-capacity links from the APs to the CPU is more challenging in cell-free Massive MIMO systems, as due to the large number of antennas at the APs, a large number of quantized signals should be sent back to the CPU. In this paper, following the terminology in \cite{marzetta_free16,ouricc1}, we refer to these links as \textit{backhaul links}.
The implementation of cell-free Massive MIMO with limited backhaul links is a more crucial challenge on the uplink, as the limited backhaul links send the quantized version of the received signals at the APs to the CPU, which introduces additional self-interference to the signals at the CPU. The total data rate required to transmit these quantized signals with sufficient precision to avoid performance loss is several times the total user data rate supported by those signals. In the C-RAN literature this has been estimated as 20-50 times the corresponding data rate \cite{CRAN_china}, implemented
using the common public radio interface (CPRI) standard \cite{CRAN_CPRI}, typically over optical fiber.
The assumption of infinite backhaul in \cite{marzetta_free16}
is not realistic in practice. It is reasonable to assume, however, that the backhaul network
will carry quantized signals, at least in the uplink direction, and
that this will affect the network performance. The current paper considers optimum uniform quantization. J. Max in \cite{max_quantization} developed
an algorithm to solve the problem of minimizing the mean-squared distortion (or mean-squared error (MSE)). In addition, P. Zillmann in \cite{Zillmann} studied the problem of minimising the MSE of the uniform quantizer exploiting \textit{the Bussgang decomposition} \cite{Buss_rep}. Note that \textit{the Max algorithm} and the scheme in \cite{Zillmann} provide the same signal-to-distortion-plus-noise ratio (SDNR). In this paper, we exploit both the Max algorithm and the Bussgang decomposition to model the optimal uniform quantization. We show that with linear detection and the exploiting optimal uniform quantization, only a few quantization bits is enough to closely approach the performance of the system with perfect backhaul links. Finally, we present the performance comparison between different linear receivers.
\vspace{-.2cm}
\section{SYSTEM MODEL}
We consider uplink transmission of a cell-free Massive MIMO system with $M$ APs and $K$ single-antenna users randomly distributed in a large area. Moreover, we assume each AP has $N$ antennas. The channel coefficient vector between the $k$th user and the $m$th AP, $\textbf{g}_{mk} \in \mathbb{C}^{N\times 1}$, is modeled as
$	
\textbf{g}_{mk}=\sqrt{\beta_{mk}}\textbf{h}_{mk},
$
where $\beta_{mk}$ denotes the large-scale fading and $\textbf{h}_{mk}\sim  \mathcal{CN}(0,\bf{I}_N)$ is a complex Gaussian random vector with covariance matrix $\bf{I}_N$ which represents the small-scale fading \cite{marzetta_free16}.
All pilot sequences transmitted by the $K$ users in the channel estimation phase are collected in a matrix $\bPhi \in \mathbb{C}^{\tau_p\times K}$, where $\tau_p$ is the length of the pilot sequence for each user and the $k$th column, $\pmb{\phi}_k$, represents the pilot sequence used for the $k${th} user. After performing a de-spreading operation, the MMSE estimate of the channel coefficient between the $k$th user and the $m$th AP is given by
\vspace{-.2cm}
\begin{IEEEeqnarray}{rCl}
	\!\!\!\!\!\!\hat{\textbf{g}}_{mk}\!=\!c_{mk}\!\Big(\!\!\sqrt{\tau_p p_p}\textbf{g}_{mk}\!\!+\!\!\sqrt{\tau_p p_p}\sum_{k^\prime\ne k}^{K}\textbf{g}_{mk^\prime}\pmb{\phi}_{k^\prime}^H\pmb{\phi}_{k}\!\!+\!\!\bomega_{p,m}\pmb{\phi}_k\!\Big),
	\label{ghat}
\end{IEEEeqnarray}where $\bomega_{p,m}$ denotes the noise vector at the $m$th antenna whose elements are independent identically distributed (i.i.d.) $\mathcal{CN}(0,1)$, $p_p$ represents the normalized signal-to-noise ratio (SNR) of each pilot sequence (which we define in Section VI), and $c_{mk}$ is given by
$
c _{mk}=\frac{\sqrt{\tau_p p_p}\beta_{mk}}{\tau_p p_p\sum_{k^\prime=1}^{K}\beta_{mk^\prime}|\pmb{\phi}_{k^\prime}^H{\pmb{\phi}}_k|^2+1}
$ \cite{marzetta_free16}.
Note that, as in \cite{marzetta_free16}, we assume that the large-scale fading,
$\beta_{mk}$, is known.
The estimated channels in (\ref{ghat}) are used by the APs to design the receiver coefficients and determine power allocations.
Using the analysis in \cite{marzetta_free16}, the mean-square of the $n$th component of the estimated channel is given by $\gamma_{mk}\triangleq  \mathbb{E}\left\{\left|\left[\hat{\textbf{g}}_{mk}\right]_n\right|^2\right\}=\sqrt{\tau_p p_p}\beta_{mk}c_{mk}$.
Next, we consider the uplink data transmission, where all users send their signals to the APs.
The transmitted signal from the $k$th user is represented by
$
x_k= \sqrt{q_k}s_k,
$
where $s_k$ ($\mathbb{E}\{|s_{k}|^2\} = 1$) and $q_k$ denotes the transmitted symbol and the transmit power from the \textit{k}th user, respectively. 
The $N\times 1$ received signal at the $m$th AP from all users is given by 
\vspace{-.2cm}
\begin{equation}
\textbf{y}_m= \sqrt{\rho}\sum_{k=1}^{K}\textbf{g}_{mk}\sqrt{q_k}s_k+\textbf{n}_m, 
\label{ym}
\end{equation}
where each element of $\textbf{n}_m \in \mathbb{C}^{N\times 1}$, $n_{n,m}\sim \mathcal{CN}(0,1)$ is the noise at the $m$th AP.\vspace{-.15cm}
\section{Optimal uniform quantization model}\label{section_quantization}
In this section, we study optimal uniform quantization. Note that J. Max in \cite{max_quantization} developed an algorithm to define the necessary conditions to minimize the distortion of the quantizer \cite{Gallagerbook}. In addition, the Bussgang decomposition \cite{Buss_rep} is used in this paper, enabling us to exploit the scheme proposed by P. Zillmann in \cite{Zillmann} to model the quantization and hence find the optimum step-size of the quantizer by maximizing the SDNR. Note that the Max algorithm and the scheme based on the Bussgang decomposition in \cite{Zillmann} result in the same SDNR. The main difference between them is that using Bussgang decomposition, the output of the quantizer can be represented by a scalar multiple of the input plus an uncorrelated distortion \cite{Buss_rep,Zillmann} whereas exploiting the Max algorithm, the quantization distortion and the output of the quantizer are uncorrelated \cite{max_quantization}. The details of the optimal uniform quantization models are provided in the following subsections.
\vspace{-.3cm}
\subsection{Optimal Uniform Quantization with Bussgang Theorem}
Based on the Bussgang decomposition \cite{Buss_rep}, the output of a quantizer can be represented by a scalar multiple of the input plus uncorrelated distortion as follows \cite{Zillmann,ourvtc18,ourjournal2}:
$
\mathcal{Q} (z) = h(z)=az+n_d, ~\forall k, 
$
where $a$ is a constant, $n_d$ refers to the distortion noise which is uncorrelated with the input of the quantizer, $z$. The term $a$ is given by 
$
a=\frac{\mathbb{E}\left\{zh(z)\right\}}{\mathbb{E}\{z^2\}}=\frac{1}{p_z}\int_{\mathcal{Z}}zh(z)f_z(z)d~z ,
$
where $p_z=\mathbb{E}\{|z|^2\}=\mathbb{E}\{z^2\}$ denotes the power of $z$ and we drop absolute value as $z$ is a real number, and $f_z(z)$ represents the probability distribution function of $z$. Moreover, we define a second parameter 
$
b=\frac{\mathbb{E}\left\{h^2(z)\right\}}{\mathbb{E}\{z^2\}}=\frac{1}{p_z}\int_{\mathcal{Z}}h^2(z)f_z(z)d~z
$ \cite{Zillmann,ourvtc18,ourjournal2}. We aim to maximize the SDNR, which is defined as follows:
$
\text{SDNR}=\frac{\mathbb{E}\left\{(az)^2\right\}}{\mathbb{E}\{n_d^2\}}=\frac{a^2}{b-a^2},
$
where $\mathbb{E}\left\{az^2\right\}=a^2p_z$, and $\mathbb{E}\{n_d^2\}=p_{n_d}=(b-a^2)p_z$.
Note that in practice, we divide the input by its standard deviation, and multiply the output by the same factor. Hence, by introducing a new variable $\tilde{z}=\frac{z}{\sqrt{p_z}}$, we have 
\begin{equation}
\mathcal{Q}(z)=\sqrt{p_z}\mathcal{Q}(\tilde{z})=\tilde{a}\sqrt{p_z}\tilde{z}+\sqrt{p_z}\tilde{n}_d=\tilde{a}z+\sqrt{p_z}\tilde{n}_d,
\label{sigmamultiplyq}
\end{equation}
where $\tilde{a}$ is a constant value which depends only on the number of quantization bits, $\alpha$, and the quantizer step-size. Hence, the optimal step-size of the quantizer can be obtained by solving the following maximization problem:
$
\Delta_\text{opt}=\argmax_{\Delta}~{\text{SDNR}}.
$
where $\Delta$ is the step-size of the quantizer. In \cite{ourvtc18,ourjournal2}, we solve (by numerical optimization) the maximization problem and the resulting $\tilde{a}$ are summarized in Table \ref{tablezillmann}.
\begin{table}[!t]
\centering 
\caption{The optimal step-size and distortion power of a uniform quantizer \textit{with and without the Bussgang decomposition}. } 
\vspace{-.3cm}
\label{tabledelta} \label{tablezillmann}
 \begin{tabular}{c c c c c}
 \hline
\\
$\alpha$  & $\!\!\!\Delta_{\text{opt}}$ & $\!\!\!\!\!\!\!\!\!\sigma_{\tilde{n}_d}^2=\tilde{b}-\tilde{a}^2=\sigma_{\tilde{e},B}^2$ & $\!\!\!\!\tilde{a}$ & $\!\!\!\sigma_{\tilde{n}_d}^2=\sigma_{\tilde{e}}^2$   \\ [.1ex] 
 \hline\hline
 \vspace{.01cm}
{1} & 1.596 &  0.2313 &  0.6366 & 0.3634\cite{max_quantization}\\[.02ex] 
 \hline
 \vspace{.01cm}
{2} &0.9957 &  0.10472 & 0.88115 & 0.1188\cite{max_quantization}\\ [.02ex] 
 \hline
  \vspace{.01cm}
{3} &0.586 &  0.036037  &0.96256 & 0.03744\cite{max_quantization}\\ [.02ex] 
 \hline
  \vspace{.01cm}
{4} &0.3352 & 0.011409  & 0.98845 & 0.01154\cite{max_quantization}\\ [.02ex] 
 \hline
  \vspace{.01cm}
{5} &0.1881  & 0.003482  & 0.996505 &0.00349\cite{max_quantization}\\ [.02ex] 
 \hline	
   \vspace{.01cm}
{6} &0.1041  & 0.0010389  & 0.99896 &-\\ [.02ex] 
 \hline	
   \vspace{.01cm}
{7} &0.0568 & 0.0003042  & 0.99969 &-\\ [.02ex] 
 \hline	
   \vspace{.01cm}
{8} &0.0307  & 0.0000876  & 0.999912&-\\ [.02ex] 
 \hline	
   \vspace{.01cm}
{9} &0.0165 &0.0000249 & 0.999975 &- \\ [.02ex] 
 \hline	
 \end{tabular}
\end{table} 
\vspace{-.2cm}
\subsection{Max Algorithm for Optimal Uniform Quantization}
Based on the analysis provided by J. Max in \cite{max_quantization}, the linear quantization can be modeled as:
\begin{equation}
\mathcal{Q} (z) = h(z)=z+n_d, ~\forall k, 
\label{zz}
\end{equation}
where the output of the quantizer and the distortion are uncorrelated \cite{max_quantization,prokis_com,mezghani_WSA16}. For this case, to calculate the variance of the quantization error, we exploit the following schemes:
\vspace{-.15cm}
\begin{equation}
\sigma_{\tilde{n}_d}^2=\left\{
\begin{array}{rl}
&\!\!\!\!\!\!\!\sigma_{\tilde{e}}^2, ~~~~~~~\text{obtained}~\text{in} $\cite{max_quantization}$ ,~~~~\alpha \le 5,\\
&
\!\!\!\!\!\!\!\tilde{a} (1-\tilde{a}),$~~\cite{mezghani_isit12}$, ~~~~~~~~~~~~~~~~\!\alpha\ge 6,
\end{array} \right.
\label{fr}
\end{equation}where $\alpha$ denotes the number of quantization bits. 
\vspace{-.25cm}
\section{Limited Backhaul}
In this section, we present the performance analysis for the limited-backhaul cell-free Massive MIMO system.
The $m$th AP quantizes the estimated channels, $\hat{\textbf{g}}_{mk}$, $\forall k$, and the received signal, $\textbf{y}_m$, using the optimal uniform quantization, and forwards the quantized channel and the quantized signal in each symbol duration to the CPU. In the following subsections, we exploit the Bussgang decomposition \cite{Buss_rep,Zillmann} and the Max algorithm \cite{max_quantization} to quantize the received signal and the estimated channel, respectively. These enable us to exploit the scheme in \cite{hien_book} to derive the SINR of the limited-backhaul cell-free Massive MIMO system.
\vspace{-.25cm}
\subsection{Quantization of the Received Signal}
Using the Bussgang decomposition \cite{Buss_rep,Zillmann}, the quantized signal can be obtained as:
\begin{equation}
[\check{\textbf{y}}_{m}]_n= \tilde{a}[\textbf{y}_{m}]_n+[\textbf{e}_{m}^y]_n ~\forall m ~\& ~\forall n. 
\end{equation}
Exploiting the analysis in Section \ref{section_quantization}, variance of the quantization error is given by $\sigma_{[\textbf{e}_{m}^y]_n}^2=\sigma_{[\tilde{\textbf{e}}_{m}^y]_n}^2\mathbb{E}\left\{\left|\left[\textbf{y}_{m}\right]_n\right|^2\right\}$. Hence, we have
\vspace{-.29cm}
\begin{IEEEeqnarray}{rCl}\label{sigma_y_expand_case1}
\sigma_{[\textbf{e}_{m}^y]_n}^2&=&\sigma_{[\tilde{\textbf{e}}_{m}^y]_n}^2\left(\rho\sum_{k^\prime=1}^K\beta_{mk^\prime}q_{k^\prime}+1\right)\nonumber \\
&=&\sigma_{\tilde{e}^y}^2\left(\rho\sum_{k^\prime=1}^K\beta_{mk^\prime}q_{k^\prime}+1\right),\forall m,n,
\vspace{-.1cm}
\end{IEEEeqnarray}
where $\sigma_{\tilde{\textbf{e}}_{m}^y}^2$ is variance of the quantization error with unit variance input for the given number of quantization bits. Moreover, in the second equality in (\ref{sigma_y_expand_case1}) we used the same number of bits in all APs and all antennas to quantize the received signal and hence $\sigma_{[\tilde{\textbf{e}}_{m}^y]_n}^2=\sigma_{\tilde{e}^y}^2=\sigma_{\tilde{e},B}^2, \forall m,n$. The optimal values of $\sigma_{\tilde{e},B}^2$ for different numbers of quantization bits are given in Table \ref{tablezillmann}.
\vspace{-.16cm}
\begin{remark}\label{remark_y_buss}
Using the Bussgang decomposition, the quantizer input is uncorrelated with the quantization error. This implies that:
$
\mathbb{E}\left\{ {\textbf{y}}_{m}^H \textbf{e}_{m}^y \right\}=0,~\forall k.
$
\end{remark}
\vspace{-.4cm}
\subsection{Quantization of the Estimated Channel}
We quantize the estimated channel with the optimal quantizer obtained using the Max algorithm \cite{max_quantization} as follows:
\vspace{-.1cm}
\begin{equation}
[\check{\textbf{g}}_{mk}]_n\!=\![\hat{\textbf{g}}_{mk}]_n\!+\![\textbf{e}_{mk}^{g}]_n,\forall k~\&~\forall n.
\end{equation}
Using the analysis in Section IV, the variance of the quantization error is obtained as $\sigma_{[\textbf{e}_{mk}^g]_n}^2=\sigma_{[\tilde{\textbf{e}}_{mk}^g]_n}^2\mathbb{E}\left\{\left[\hat{\textbf{g}}_{mk}]_n\right|^2\right\}$, which results in
\begin{equation}\label{sigma_g_expand}
\sigma_{[\textbf{e}_{mk}^g]_n}^2=\sigma_{[\tilde{\textbf{e}}_{mk}^g]_n}^2\gamma_{mk}=\sigma_{\tilde{e}^g}^2\gamma_{mk},~\forall m,k,n,
\end{equation}
where for simplicity we use the same number of bits in all APs to quantize the estimated channel.
\vspace{-.15cm}
\begin{remark}\label{remark_g_opt}
Based on \cite{max_quantization,mezghani_WSA16,prokis_com}, the quantizer output is assumed to be uncorrelated with the quantization error. Hence, we have
$
\mathbb{E}\left\{\check{\textbf{g}}_{mk}^H \textbf{e}_{mk}^{g} \right\}=0,~\forall k.
$
\end{remark}
\vspace{-.15cm}
\begin{remark}\label{remark_quant_mean}
If the probability density function of input of the quantizer is even, and exploiting the symmetrical quantizer, the quantization error has zero mean \cite{max_quantization}. Hence, we have:
$
\mathbb{E}\left\{ \textbf{e}_{mk}^g\right\}=0$ and $\mathbb{E}\left\{ \textbf{e}_{m}^y
\right\}=0.
$
\end{remark}
\vspace{-.1cm}
\subsection{Data Detection}
Let $\check{\textbf{V}} \in \mathbb{C}^{MN\times K}$ be linear detector matrix depending on the side information at the receiver $\check{\textbf{g}}_{mk}, \forall m,k$. We let $\check{\textbf{v}}_k=\left[\check{\textbf{v}}_{1k}^T\cdots\check{\textbf{v}}_{Mk}^T\right]^T$ refer to the $k$th column of the detector matrix $\check{\textbf{V}}$, and $\check{\textbf{v}}_{mk} \in \mathbb{C}^{N}$. The estimate of the transmitted data $s_k$ is given by
\begin{IEEEeqnarray}{rCl}
\check{s}_k=\check{\textbf{v}}_k^H \left[\check{\textbf{y}}_1^T\cdots\check{\textbf{y}}_M^T\right].
\end{IEEEeqnarray}
Next, the received signal for the $k$th user after using the detector at the CPU is given by
\begin{IEEEeqnarray}{rCl}
&&\!\!\!\!\!\!\!r_k \!\!=\!\!\sum_{m=1}^{M}\check{\textbf{v}}_{mk}^H\check{\textbf{y}}_{m}=\sum_{m=1}^M\check{\textbf{g}}_{mk}^H\left(\tilde{a}\textbf{y}_m+\textbf{e}_{m}^y\right)
\\
&=&\sum_{m=1}^M\check{\textbf{v}}_{mk}^H\left(\tilde{a}\sqrt{\rho}\sum_{k=1}^{K}\textbf{g}_{mk}\sqrt{q_k}s_k+a\textbf{n}_m+\textbf{e}_{m}^y\right)\nonumber \\
&=&\sum_{m=1}^M\check{\textbf{v}}_{mk}^H\left(\tilde{a}\sqrt{\rho}\sum_{k=1}^{K}\left(\check{\textbf{g}}-\textbf{e}_{mk}^g-\tilde{\textbf{g}}_{mk}\right)\sqrt{q_k}s_k+a\textbf{n}_m+\textbf{e}_{m}^y\right)\nonumber\\
&=&
\tilde{a}\underbrace{\sqrt{\rho q_k}\sum_{m=1}^{M}\!\check{\textbf{v}}_{mk}^H\check{\textbf{g}}_{mk}}_{A_1}s_k+\tilde{a}\underbrace{\sqrt{\rho}\sum_{k^\prime\neq k}^{K}\sqrt{q_{k^\prime}}\sum_{m=1}^{M}\check{\textbf{v}}_{mk}^H\check{\textbf{g}}_{mk^\prime}}_{A_2}s_{k^\prime}\nonumber
\\
&+&\tilde{a}\underbrace{\sum_{m=1}^{M}\check{\textbf{v}}_{mk}^H\textbf{n}_m}_{A_3}+\underbrace{\sum_{m=1}^{M}\check{\textbf{v}}_{mk}^H\textbf{e}_{m}^y}_{A_4}-\nonumber\\
&\!\!\!\!\!\!\!\!\!&\!\!\!\!\!\!\!\!\tilde{a}\!\underbrace{\sqrt{\rho}\!\sum_{m=1}^{M}\!\check{\textbf{v}}_{mk}^H\sum_{k^\prime=1}^{K}\!\!\sqrt{q_{k^\prime}}\textbf{e}_{mk^\prime}^g}_{A_5}s_{k^\prime}\!\!-\!\!\tilde{a}\underbrace{\sqrt{\rho}\!\sum_{m=1}^{M}\!\check{\textbf{v}}_{mk}^H\!\!\sum_{k^\prime=1}^{K}\!\sqrt{q_{k^\prime}}\tilde{\textbf{g}}_{mk}}_{A_6}s_{k^\prime}.\nonumber
 \label{rkcase1_limited}
\end{IEEEeqnarray}
\begin{lemma}\label{lemma_all_lb_ap}
Terms $A_1$, $A_2$, $A_3$, $A_4$, $A_5$ and $A_6$ are mutually uncorrelated.
\end{lemma}
\textit{Proof:} Please refer to Appendix.~~~~~~~~~~~~~~~~~~~~~~~~~~~~~~~~~~$\blacksquare$
Using Lemma \ref{lemma_all_lb_ap} and analysis in \cite[Table 2.3]{hien_book}, the SINR of the $k$th user is obtained by the following theorem.
\begin{theorem}\label{theorem_rkcase1_limited}
The ergodic achievable rate of the $k$th user in cell-free Massive MIMO for the case when the APs estimate the channel and send back the quantized version of the estimated channel and the quantized version of the received signal through the limited backhaul links is given by $R_k^{lb} =E\{\log_2(1+SINR_k^{lb}) \}$, where superscript “lb” refers to limited backhaul links, and $SINR_k^{lb}$ is given by (\ref{sinr_propoition_rkcase1_limited})
\vspace{-.2cm}
\begin{IEEEeqnarray}{rCl}\label{sinr_propoition_rkcase1_limited}
\!\!\!\!\!\!\text{SINR}_k^{\text{lb}}\left(\check{\textbf{v}}\right)=\frac{\rho q_k \check{\textbf{v}}_k^H\check{\textbf{g}}_k\check{\textbf{g}}_k^H\check{\textbf{v}}_k}
{\check{\textbf{v}}_k^H\left(\rho \sum_{k^\prime\neq k}^{K}q_{k^\prime}\check{\textbf{g}}_{k^\prime}\check{\textbf{g}}_{k^\prime}^H+\textbf{R}^{\text{lb}}\right)\check{\textbf{v}}_k}\!,~
\end{IEEEeqnarray}
where $\check{\textbf{g}}_k=\left[\check{\textbf{g}}_{1k}^T\cdots\check{\textbf{g}}_{Mk}^T\right]^T$, $\textbf{R}^{\text{lb}}$ is obtained as follows:
\vspace{-.4cm}
\end{theorem}
\vspace{-.4cm}
\begin{subequations}
\begin{align}
&\textbf{R}^{\text{lb}}=\rho \sum_{k^\prime=1}^Kq_{k^\prime}\textbf{W}_{k^\prime}^{\text{lb}}+\textbf{I}_{ M N}+\textbf{F}^{\text{lb}},
\label{R_lb_ap1}\\
&\textbf{W}_{k^\prime}^{\text{lb}}=\textbf{S}_{k^\prime}^{\text{lb}}-\textbf{T}_{k^\prime}^{\text{lb}}, \textbf{F}^{\text{lb}}=\dfrac{\sigma_{\tilde{e}^y}^2}{\tilde{a}^2}\textbf{I}_{MN},
\label{R_lb_ap2}\\
&\!\!\!\textbf{S}_{k^\prime}^{\text{lb}}
\!\!=\!\!
\left(\!\!\frac{\sigma_{\tilde{e}^y}^2}{\tilde{a}^2}\!\!+\!\!1\!\!\right)\!\text{diag}\left(\text{rep}\left(\!\beta_{1k^\prime}
,N\!\right) 
\!\cdots\!
\text{rep}\left(\beta_{Mk^\prime}
,N\!\right)\right),
\label{R_lb_ap3}\\
&\!\!\!\textbf{T}_{k^\prime}^{\text{lb}}\!=\!\left(\!1\!-\!\sigma_{\tilde{e}^g}^2\!\right)\text{diag}\left(\text{rep}\left(\gamma_{1k^\prime},N\right)\cdots\text{rep}\left(\gamma_{Mk^\prime},N\right)\right),
\label{R_lb_ap4}
\end{align}
\end{subequations}\textit{where} $\text{rep}\left(x
,N\right)=[x\cdots x]\in \mathbb{C}^{1\times N}$.
\vspace{-.01cm}
\begin{figure*}[t]
\begin{IEEEeqnarray}{rCl}
\begin{split}
\label{sinr_A1A6}
\text{SINR}_k^{\text{lb}}=\frac{
\mathbb{E}\left\{\left|A_1|\check{\textbf{g}}_{k}\right|^2\!\right\}}
{
\mathbb{E}\left\{\left|A_2|\check{\textbf{g}}_{k}\right|^2\right\}+
\mathbb{E}\left\{\left|A_3|\check{\textbf{g}}_{k}\right|^2\right\}+
\frac{1}{\tilde{a}^2}\mathbb{E}\left\{\left|A_4|\check{\textbf{g}}_{k}\right|^2\!\right\}+
\mathbb{E}\left\{\left|A_5|\check{\textbf{g}}_{k}\right|^2\right\}+
\mathbb{E}\left\{\left|A_6|\check{\textbf{g}}_{k}\right|^2\right\}
}.
\end{split}
\vspace{-.27cm}
\end{IEEEeqnarray}
\vspace{-.19cm}
\hrulefill
\end{figure*}

\textit{Proof:} Using Lemma \ref{lemma_all_lb_ap} and the analysis in \cite{hien_book}, the achievable SINR is obtained by (\ref{sinr_A1A6}) (provided at the top of next page).
It is easy to show that the achievable SINR is obtained by (\ref{sinr_propoition_rkcase1_limited}). In addition, using (\ref{sigma_y_expand_case1}) and (\ref{sigma_g_expand}), and after some mathematical manipulation, we have
\begin{IEEEeqnarray}{rCl}\label{simlyfy_a4a5a6}
&&\!\!\!\!\dfrac{1}{\tilde{a}^2}\mathbb{E}\left\{\left|A_4|\check{\textbf{g}}_{k}\right|^2\right\}+
\mathbb{E}\left\{\left|A_5|\check{\textbf{g}}_{k}\right|^2\right\}+
\mathbb{E}\left\{\left|A_6|\check{\textbf{g}}_{k}\right|^2\right\}
\nonumber \\
&=&\!\sum_{m=1}^M\left||\check{\textbf{v}}_{mk}\right||^2\sum_{k^\prime=1}^k\rho q_{k^\prime}\!\!\left[\!\beta_{mk^\prime}\left(1+\frac{\sigma_{\tilde{e}}^2}{\tilde{a}^2}
\right)
\!-\!
\gamma_{mk^\prime}\left(1-\sigma_{\tilde{e}}^2\!\right)\!\right]
\nonumber\\
&+&
\sum_{m=1}^M\!\left||\check{\textbf{v}}_{mk}\right||^2\frac{\sigma_{\tilde{e}}^2}{\tilde{a}^2}=\check{\textbf{v}}_k^H\left(\rho \sum_{k^\prime=1}^Kq_{k^\prime}\textbf{W}_{k^\prime}^{\text{lb}}+\textbf{F}^{\text{lb}}\right)\check{\textbf{v}}_k.
\end{IEEEeqnarray}
By substituting (\ref{simlyfy_a4a5a6}) into (\ref{sinr_A1A6}), it is easy to show that the closed-form SINR can be obtained as in (\ref{sinr_propoition_rkcase1_limited}), which completes the proof of Theorem \ref{theorem_rkcase1_limited}. ~~~~~~~~~~~~~~~~~~~~~~~~~~~~~~~~~~~~~~~~~~~~$\blacksquare$

Note that the linear detector is given by
\begin{equation}
\check{\textbf{V}}=\left\{
\begin{array}{rl}
&\!\!\!\!\!\!\check{\textbf{G}} ,~~~~~~~~~~~~~~~~~~~~~~~~~~~~~~~~~~~~~~~~\text{MRC}\\
&\!\!\!\!\!\!\left(\check{\textbf{G}}\check{\textbf{G}}^H\right)^{-1}\check{\textbf{G}}, ~~~~~~~~~~~~~~~~~~~~~~~~~~\text{ZF}\\
&\!\!\!\!\!\!\left(\tilde{a}^2\rho \sum\limits_{k^\prime=1}^Kq_{k^\prime}\check{\textbf{g}}_{k^\prime}\check{\textbf{g}}_{k^\prime}^H+\textbf{R}^{\text{lb}}\right)^{-1}\check{\textbf{G}},~ \text{MMSE}
\end{array} \right.
\label{fr}
\end{equation}
where $\check{\textbf{G}}=\left[\check{\textbf{g}}_{1}\cdots\check{\textbf{g}}_{K}\right]$.
\vspace{-.3cm}
\subsection{The required capacity for backhaul links}
Let us assume the length of the uplink data is $
\tau_f = \tau_c - \tau_p,
$
where $\tau_c$ denotes the number of samples for each coherence interval. The required number of bits for each AP to quantize the estimated channel and the uplink data during each coherence interval is $2\alpha\times(NK+N\tau_f)$, where again $\alpha$ is the number of quantization bits at each AP to quantize the estimated channel and the received signal. Finally $R_{\text{bh},m}$ represents the backhaul rate of cell-free Massive MIMO and is given by
\vspace{-.2cm}
\begin{equation}
R_{\text{bh,m}}=\dfrac{2\alpha\left(NK+N\tau_f\right)}{T_c}
\label{rbh_conf}
\end{equation}
where $T_c$ (in sec.) refers to coherence time. 
\vspace{-.3cm}
\section{Numerical Results and Discussion}
In this section, we provide numerical results to evaluate the performance of cell-free massive MIMO with different schemes. A cell-free Massive MIMO system with $M$ APs and $K$ single-antenna users is considered in a $D \times D$ simulation area, where both APs and users are uniformly distributed in random locations. 
To model the channel coefficients between users and APs, the coefficient $\beta_{mk}$ is given by 
$
\beta_{mk} = \text{PL}_{mk}. 10^{\frac{\sigma_{sh}z_{mk}}{10}}
\label{beta1}
$
where $\text{PL}_{mk}$ is the path loss from the $k$th user to the $m$th AP, and $10^{\frac{\sigma_{sh}~z_{mk}}{10}}$ denotes the shadow fading with standard deviation
$\sigma_{sh}$, and $z_{mk} \sim  \mathcal{N}(0,1)$ \cite{marzetta_free16}. The noise power is given by
$
P_n=\text{BW}k_BT_0W,
$
where $\text{BW}=20$ MHz denotes the bandwidth, $k_B = 1.381 \times 10^{-23}$ represents the Boltzmann constant, and $T_0 = 290$ (Kelvin) denotes the noise temperature. Moreover, $W=9$dB, and denotes the noise figure \cite{marzetta_free16}. It is assumed that that $\bar{P}_p$ and $\bar{\rho}$ denote the pilot sequence and the uplink data powers, respectively, where $P_p=\frac{\bar{P}_p}{P_n}$ and $\rho=\frac{\bar{\rho}}{P_n}$. In simulations, we set $\bar{P}_p=100$ mW and $\bar{\rho}=100$ mW.
Similar to \cite{marzetta_free16}, we suppose the simulation area is wrapped around at the edges, and hence can simulate an area without boundaries. We evaluate the rate of the system over 300 random realizations of the locations of APs, users and shadowing.
\begin{figure}[t!]
\vspace{-.223in}
\center
\includegraphics[width=77mm]{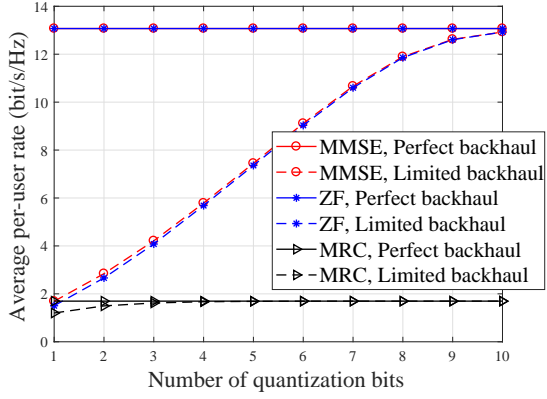}
\vspace{-.1in}
\caption{The average uplink per-user rate versus the number of quantization bits with $M=5$, $N=20$, $K=40$, $\tau_p=40$, and $D=1$ km.}
\label{conf_m5_n20_k40_t40_pp1_d1}
\end{figure}
First, the average per-user rate performance of different cases are investigated. 
Fig. \ref{conf_m5_n20_k40_t40_pp1_d1} presents the sum rate performance of the cell-free Massive MIMO system with $M=5$ APs and $K=40$ users, and $D=1$ km. Moreover, we consider orthogonal pilot sequences, i.e., $\tau_p=K$, and assume each AP is equipped with $N=20$ antennas.
As the figure demonstrates, for MRC to closely approach the performance of perfect backhaul links, we need to set $\alpha\ge 4$. However, as ZF and MMSE are more sensitive to quantization error, we need to set $\alpha\ge 9$ to approach the performance of perfect backhaul links.
\begin{figure}[t!]
\center
\vspace{-.224in}
\includegraphics[width=77mm]{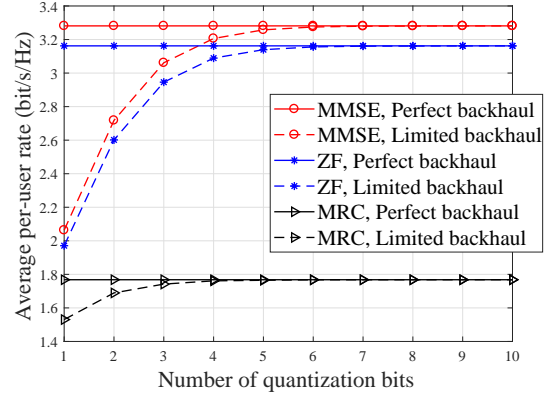}
\vspace{-.1in}
\caption{The average uplink per-user rate versus the number of quantization bits with $M=100$, $N=2$, $K=40$, $\tau_p=30$, and $D=1$ km.}
\label{conf_m100_n2_k40_t30_pp1_d1_rbh}
\end{figure}
Next, the average sum rate performance of the cell-free Massive MIMO system with random pilot assignment and more APs is investigated. Fig. \ref{conf_m100_n2_k40_t30_pp1_d1_rbh} shows the average sum rate with different linear receivers and $M=100$, $N=2$, $K=40$, $D=1$ km, and $\tau_p=30$. 
As the figure demonstrates, the performance of the system with limited backhaul links reaches the performance of the system with perfect backhaul links with fewer quantization bits compared to Fig. \ref{conf_m5_n20_k40_t40_pp1_d1}. This is the case for all linear receivers, and can be observed in Fig. \ref{conf_m5_n20_k40_t40_pp1_d1}.
\begin{figure}[t!]
\center
\vspace{-.18in}
\includegraphics[width=77mm]{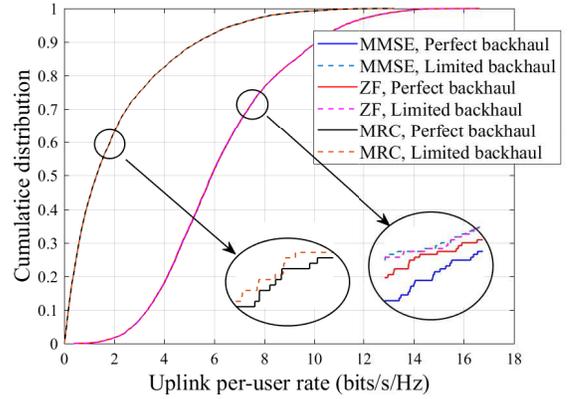}
\vspace{-.1in}
\caption{The cumulative distribution of the uplink user rate for the case of $M=10$, $N=25$, $K=40$, $\tau_p=40$, and $D=1$ km, and $\alpha=10$.}
\label{conf_m10_n25_k40_t40_pp1_d1_b10}
\end{figure}
\begin{figure}[t!]
\center
\vspace{-.24in}
\includegraphics[width=77mm]{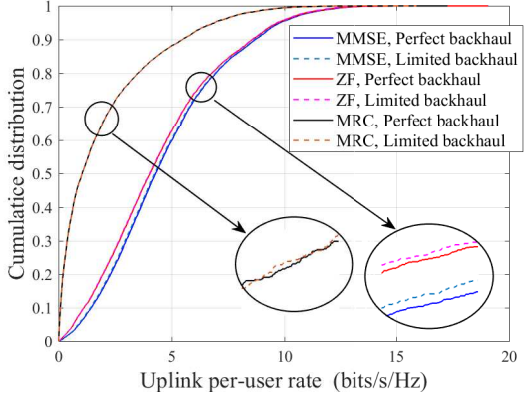}
\vspace{-.1in}
\caption{The cumulative distribution of the uplink user rate for the case of $M=10$, $N=25$, $K=40$, $\tau_p=30$, $D=1$ km, and $\alpha=8$.}
\label{conf_cdf_m10_n25_k40_t30_pp1_d1}
\end{figure}
Next, we investigate the cumulative distribution function of per-user uplink rate with different distributions of the total number of service antennas. In Fig. \ref{conf_m10_n25_k40_t40_pp1_d1_b10}, the cumulative distribution of per-user uplink rates of cell-free Massive MIMO is investigated while we set $M=10$, $N=25$, $K=40$, $\tau_p=40$, and $D=1$ km. Moreover, we assume $\alpha=10$ bits for quantization. Similar to \cite{marzetta_free16}, $T_c=1$ ms denotes the coherence time and $\tau_c=200$ is the number of samples for each coherence interval. Hence, using (\ref{rbh_conf}), the required capacity for backhaul links for the network set-up in Fig. \ref{conf_m10_n25_k40_t40_pp1_d1_b10} can be calculated as
\vspace{-.12cm}
\begin{equation}
R_{\text{bh}}^{\text{req}}=\frac{2\alpha\left(NK+N\tau_f\right)}{T_c}=100~\text{Mbits}/\text{s}.
\label{rbh_req1}
\end{equation}
Note that based on the model in \cite{fettwis_globe11}, it is reasonably practical to consider $R_{\text{bh}}^{\text{req}}=100$ \text{Mbits}/s for the capacity of wireless microwave backhaul links. 
In addition, Fig. \ref{conf_m10_n25_k40_t40_pp1_d1_b10} reveals that wireless backhaul links with a capacity of 100 Mbits/s is enough to approach the performance of perfect backhaul links. Moreover, it can be observed that performance of the ZF receiver is almost as good as the performance of the MMSE receiver. Fig. \ref{conf_cdf_m10_n25_k40_t30_pp1_d1}  investigates the performance comparison with $K=40$, $\alpha=8$ and random pilot assignment with $\tau_p=30$, $M=10$ and $N=25$. As the figure shows, the per-user uplink rate of the cell-free Massive MIMO system with limited backhaul links and $\alpha=8$ quantization bits is very close to the performance of the cell-free Massive MIMO system with perfect backhaul links.
\vspace{-.245cm}
\section{Conclusions}
The performance of the cell-free Massive MIMO with limited backhaul links has been presented. The CPU uses the quantized channel estimates and linear processing schemes to detect the desired signals from the quantized data signals. The Max algorithm has been exploited to model the optimal uniform quantization. Moreover, we used the Bussgang decomposition, which enables us to find a linear relationship between the input of the quantizer and the quantization noise. Achievable rates with different linear receivers have been determined. Numerical results have been provided to demonstrate a comparison between the cases of limited backhaul and perfect backhaul links, which reveals that the performance of limited-backhaul cell-free Massive MIMO is close to that of the ideal system with perfect backhaul links.
\vspace{-.12cm}
\section*{Appendix: Proof of Lemma \ref{lemma_all_lb_ap}} 
\vspace{-.05in}
In the following, we show that terms $A_1$, $A_2$, $A_3$, $A_4$, $A_5$, and $A_6$ are pairwise uncorrelated for the MRC case. The proof for ZF and MMSE follows the same steps and is omitted due to space limit.
\begin{itemize}
\item[1.] \label{propa1a5} Using Remark \ref{remark_g_opt}, terms $A_1$ and $A_5$ are uncorrelated.
\item[2.]\label{propa1a4} The following equation shows that terms $A_1$ and $A_4$ are uncorrelated;
\begin{IEEEeqnarray}{rCl}
\!\!\!\!\!\!\!&\!\!\!\!\!\!\!&\!\!\!\!\!\!\!\left\{A_1^\ast A_4 \right\}=
\mathbb{E}\Bigg\{\!
 \left(\!\tilde{a}\sqrt{\rho q_k}\sum_{m=1}^{M}\!\check{\textbf{g}}_{mk}^H
 \check{\textbf{g}}_{mk}s_k\right)^H
 \\
 &&\!\!\!\!\!\!\!\left(\sum_{m=1}^{M}\check{\textbf{g}}_{mk}^H\textbf{e}_{m}^y\!\right)
\!\Bigg\} =M\tilde{a}\sqrt{\rho q_k} \mathbb{E}\left\{ ||\check{\textbf{g}}_k||^2\check{\textbf{g}}_k^H\textbf{e}^ys_k^*\right\}=0,\nonumber
\end{IEEEeqnarray}
where $\textbf{e}^y=\left[\textbf{e}_{1}^T\cdots\textbf{e}_{M}^T\right]^T$, and the second equality is due to the following facts:
$
\mathbb{E}\left\{
\check{\textbf{g}}_{k}^H s_k 
\right\}= \underline{\bold{0}}, \mathbb{E}\left\{
\check{\textbf{g}}_{k}^H \textbf{e}^y
\right\}=0,
\mathbb{E}\left\{
\textbf{e}^y s_k
\right\}=\underline{\bold{0}},
$ where $\underline{\bold{0}}=[0\cdots 0]^T \in \mathbb{C}^{MN\times 1}$.
\item[3.] \label{propa4a5} Show that terms $A_4$ and $A_5$ are uncorrelated.
\begin{IEEEeqnarray}{rCl}
\mathbb{E}\left\{A_4^\ast A_5 \right\}&=&\mathbb{E}\Bigg\{
 \left(\sum_{m=1}^{M}\check{\textbf{g}}_{mk}^H\textbf{e}_{m}^y\right)^H\\
&&
\!\left(\!\!\tilde{a}\sqrt{\rho}\sum_{m=1}^{M}\check{\textbf{g}}_{mk}^H\sum_{k^\prime=1}^{K}\sqrt{q_{k^\prime}}\textbf{e}_{mk^\prime}^gs_{k^\prime}\!\right)
\!\Bigg\}\! =\!0,\nonumber
\end{IEEEeqnarray}
where the second equality is due to the following facts:
\begin{subequations}
\begin{eqnarray}
&&\mathbb{E}\left\{
\check{\textbf{g}}_{mk}^H s_{k^ \prime}
\right\}=\bold{0},\mathbb{E}\left\{
{\textbf{e}_{mk^\prime}^g}^H s_{k^\prime}
\right\}=\bold{0},\label{fact11}\\
&&\mathbb{E}\left\{
\check{\textbf{g}}_{mk}^H \textbf{e}_{mk^\prime}^g
\right\}=0,\mathbb{E}\left\{
\check{\textbf{g}}_{mk}^H \textbf{e}_{m}^y
\right\}=0,\label{fact12}
\end{eqnarray}
\end{subequations}
where (\ref{fact11}) is due to the fact that there is no correlation between the transmitted signal $s_k$ and the quantized version of the estimated channel. Moreover, note that (\ref{fact12}) comes from Remark \ref{remark_g_opt}.
\item[4.]\label{propa2a5all}
Using Remark \ref{remark_g_opt}, terms $A_2$ and $A_5$ are uncorrelated.
\item[5.]\label{propa3a6all}
As terms $A_3$ and $A_6$ include i.i.d. Gaussian noise and i.i.d. Gaussian MMSE error, respectively, $A_2$ and $A_6$ are uncorrelated with other terms.
\end{itemize}
Using points 1, 2, 3, 4, and 5, it is easy to show that terms $A_1$, $A_2$, $A_3$, $A_4$, $A_5$ and $A_6$ are mutually uncorrelated, which completes the proof of Lemma \ref{lemma_all_lb_ap}.~~~~~~~~~~~~~~~~~~~~~~~~~~~~~~~~~~~~~$\blacksquare$
\vspace{-.4cm}
\bibliographystyle{IEEEtran}
\bibliography{cran_conf11} 
\end{document}